\documentclass[12pt,a4paper]{elsarticle_mod}
\usepackage[utf8]{inputenc}
\usepackage{natbib}
\usepackage{placeins}
\setcitestyle{square, compress}
\usepackage{mathrsfs}
\usepackage{array}
\usepackage{amssymb,amsmath,amsthm}
\usepackage{enumerate}
\usepackage{tikz}
\usepackage[official]{eurosym}
\usepackage{hyperref}
\usepackage{verbatim}
\usepackage{enumerate}
\usepackage{multirow}

\newtheorem{thm}{Theorem}[section]

\newtheorem{defn}[thm]{Definition}
\theoremstyle{definition} 

\theoremstyle{definition} 

\def\red#1{\textcolor{red}{#1}}

\def\x{\widetilde x}

\newcommand{\bb}[1]{\boldsymbol{#1}}

\theoremstyle{definition}

\begin{document}
\title{\huge Comparing district heating options under uncertainty using stochastic ordering}
\tnotetext[t1]{This paper is produced as part of the research project Managing Uncertainty in Government Modeling (MUGM) funded by the Alan Turing Institute. This paper is supported by European Union's Horizon 2020 research and innovation programme under grant agreement No 767429, project ReUseHeat.}

\author[1]{Victoria Volodina\corref{cor1}%
\fnref{fn1}}
\ead{vvolodina@turing.ac.uk}

\author[2]{Edward Wheatcroft \fnref{fn2}}
\ead{E.D.Wheatcroft@lse.ac.uk}

\author[2]{Henry Wynn \fnref{fn2}}
\ead{H.Wynn@lse.ac.uk}

\cortext[cor1]{Corresponding author}

\address[1]{The Alan Turing Institute, British Library, 96 Euston Road, London, NW1 2DB, UK}

\address[2]{London School of Economics, Houghton Street, London,  WC2A 2AZ, UK}



\noindent
\normalsize

\begin{abstract}
District heating is a network of pipes through which heat is delivered from a centralised source.  It is expected to play an important role in the decarbonisation of the energy sector in the coming years. In district heating, heat is traditionally generated through fossil fuels, often with combined heat and power (CHP) units.  However, increasingly, waste heat is being used as a low carbon alternative, either directly or, for low temperature sources, via a heat pump. The design of district heating often has competing objectives: the need for inexpensive energy and meeting low carbon targets. In addition, the planning of district heating schemes is subject to multiple sources of uncertainty such as variability in heat demand and energy prices. This paper proposes a decision support tool to analyse and compare system designs for district heating under uncertainty using stochastic ordering (dominance). Contrary to traditional uncertainty metrics that provide statistical summaries and impose total ordering, stochastic ordering is a partial ordering and operates with full probability distributions. In our analysis, we apply the orderings in the mean and dispersion to the waste heat recovery problem in Brunswick, Germany. \\
\end{abstract}


\maketitle

\vspace{5mm}

\section{Introduction}
\label{sec:Introduction}
The transition to a net-zero economy is an urgent challenge, and many countries have agreed to put in place national action plans to become carbon-neutral by 2050 or sooner \cite{EuropeGreenDeal,Paris}.  A significant aspect of the response is an increase in funding for heat decarbonisation and energy efficiency projects. In the context of tight government budgets and the uncertainty associated with the implementation of green technologies, a comprehensive handling of risk is crucial in assessing the viability of these projects. The aim of this paper is to demonstrate a new approach to dealing with uncertainty for the planning of energy infrastructure projects in the form of stochastic ordering (dominance). 

The context of this paper is waste heat recovery, in which heat from industrial and urban sources is used as a zero-carbon alternative to fossil fuels in district heating networks. In particular, we consider a district heating project based on a real system in Brunswick (Braunschweig), Germany, in which heat from a data centre will be used as an input to a district heating network for a newly constructed residential and commercial area in the city.  The system is one of four demonstrators on the ReUseHeat project \cite{lygnerud2019contracts}. The main objective of ReUseHeat is to aid \textit{replication} of projects and provide advice to other investors. Here, ``based on'' refers to the matching of variables, that is matching the model inputs and outputs to the metrics observed in the real-world. We note that the modelling results were not used in the design of the actual system. However, the model conclusions derived can provide assistance for the design of similar projects.\par

A wide range of methods have been adopted to support decisions in the energy systems domain under uncertainty. For example, \cite{Soroudi2013} pointed out the importance of uncertainty analysis in energy systems. Different approaches include the use of membership functions (fuzzy methods) and probability density functions (stochastic methods) for describing the uncertainty of input parameters. The effect of input parameters on model outputs can be quantified by simple statistical measures such as the standard deviation, mean squared error, confidence intervals and their multivariate and Bayesian counterparts. In addition, there is a recent interest in the use of scenarios to capture wider uncertainty issues \cite{Wheatcroft2019}. 

Another general approach for decision support in energy systems is optimisation frameworks (stochastic optimisation), which operate by controlling the output variability of a component or system, while keeping the output minimum (maximum) on target \cite{Moret2020}. In particular, in mean-variance (MV) and capital asset pricing models, widely used in finance and economics to assess different investment options, the aim is to achieve maximum yield and minimum volatility. Mean-variance portfolio theory has been employed for the modelling of uncertainties and risk in energy system planning \cite{ioannou2017risk, bhattacharya2012power}. These decision approaches rely on the MV rule, which states that, when two alternative prospects are considered, a rational decision-maker will choose an option with higher expected return and lower variance. However, the MV rule leads to a number of paradoxes such as the fact that it is sometimes unable to rank the two choices when a clear preference between them exists \cite{Levy2015} .

We note that the approaches described above operate with statistical summaries, which can fail to provide specific information about the distribution of outputs. We are keen to preserve the notion of robustness (uncertainty) by considering full probability distributions using stochastic ordering (dominance). Stochastic ordering is a special methodological framework, which underlines decision making under uncertainty and has been exploited in a number of fields, including robust design \cite{ cook2018optimization}, portfolio theory \cite{annaert2009performance} and signal processing \cite{tepedelenlioglu2011applications}. In particular, we note that stochastic dominance has been used previously to compare the performance of algorithms in power and energy system optimisation \cite{Chicco2017}. An extensive review of stochastic ordering is given in \cite{mosler1993stochastic}.

A stochastic ordering is a partial ordering on distributions. In this paper we consider first-order stochastic dominance (FSD) \cite{Levy2015}. Orderings provide a useful framework for comparing distributions in the following sense: suppose we are aiming to maximise the return on our investment and have two choices to make regarding the investment portfolio (Portfolio $X$ and Portfolio $Y$). We assume that there is uncertainty associated with returns on investment portfolios and we can construct probability distributions of the return in each case.  If the distribution of choice $X$ first-order stochastically dominates that of choice $Y$, for any given level of return, the probability that the profit exceeds that particular value is higher for choice $X$ than it is for choice $Y$. In other words, if we are only interested in maximising the probability of the return exceeding some chosen value, we should always choose Portfolio $X$. This seems like a sensible and intuitive way to look at decision making under uncertainty and this is our approach here. \par  

Stochastic orderings roughly divide into two classes: orderings that denote shift (in mean) and those that denote variability (dispersion). We note that the ordering does not prescribe uncertainty measures, but we claim that it provides a platform for uncertainty analysis in the following sense. Given an ordering, a suitable metric is one which is order-preserving with respect to that ordering. This means that it is a function whose expectation is ordered in the same direction as the stochastic underlying ordering. Thus both the mean and the median themselves are order-preserving with respect to first-order stochastic dominance and both the standard deviation and the Gini coefficients are ordered with respect to some well-known dispersion orderings. Following this condition, stochastic ordering allows us to extend the range of metrics for risk and uncertainty beyond the standard measures.\par



This paper is structured as follows. Section \ref{sec:District_heating} covers background regarding district heating.
Section \ref{sec:StochasticOrdering} provides an introduction to stochastic orderings. Section \ref{sec:Experiments} describes a series of simple computer experiments on a specially selected set of system designs, choices of input variations and broader scenario-based alternatives. Section \ref{sec:Results} draws conclusions based on a selection of stochastic orderings. Section \ref{sec:Discussion} consists of discussion and motivation for future work. \par

\section{Problem specification}
\label{sec:District_heating}
\subsection{District heating}
District heating is a system in which heat is produced by some centralised source and distributed via a network of insulated pipes.  It is particularly well-developed in northern Europe and Scandinavia \cite{DEA}. Historically, district heating has been powered by the burning of fossil fuels such as coal and gas.  However, more recently, it has been seen as an opportunity to decarbonise the heating sector via the use of waste heat from industry and other sources, with an increasing focus on the opportunities of recovering waste heat from low temperature, urban sources such as, for example, metro stations \cite{lygnerud2019contracts}. The prevalence of low temperature sources and their location near areas of heat demand provides an opportunity in the wider agenda of carbon reduction \cite{persson2018accessible}. \par

A major difference between high and low temperature heat recovery is that, in the latter case, the heat typically needs to be upgraded before it is suitable for use in the network.  This requires the use of a heat pump to increase the temperature to the required level.  The installation of heat pumps poses additional technical challenges due to a lack of maturity in the technology and a lack of experience in installation and maintenance. In addition, heat pumps run on electricity and this creates additional operational costs and a vulnerability to increases in the price of electricity.  This is considered to be a major risk in low temperature heat recovery \cite{lygnerud2019contracts}. \par

At present, low temperature heat recovery is not widespread.  There are a number of reasons for this, including a lack of political and commercial awareness, a lack of interest from heat 'owners', the immaturity of the technology and a lack of a legal and regulatory framework \cite{lygnerud2019contracts}. However, one of the biggest barriers, in our opinion, to the widespread rollout of low temperature waste heat recovery is a gap between the risk assessment required by financial institutions and that which is typically provided by project developers. With the methodology presented in this paper, we aim to contribute towards closing this gap. 

\subsection{Heat recovery in Brunswick}\label{section:brunswick}
\begin{figure}[ht]
\begin{center}
\includegraphics[scale=0.4]{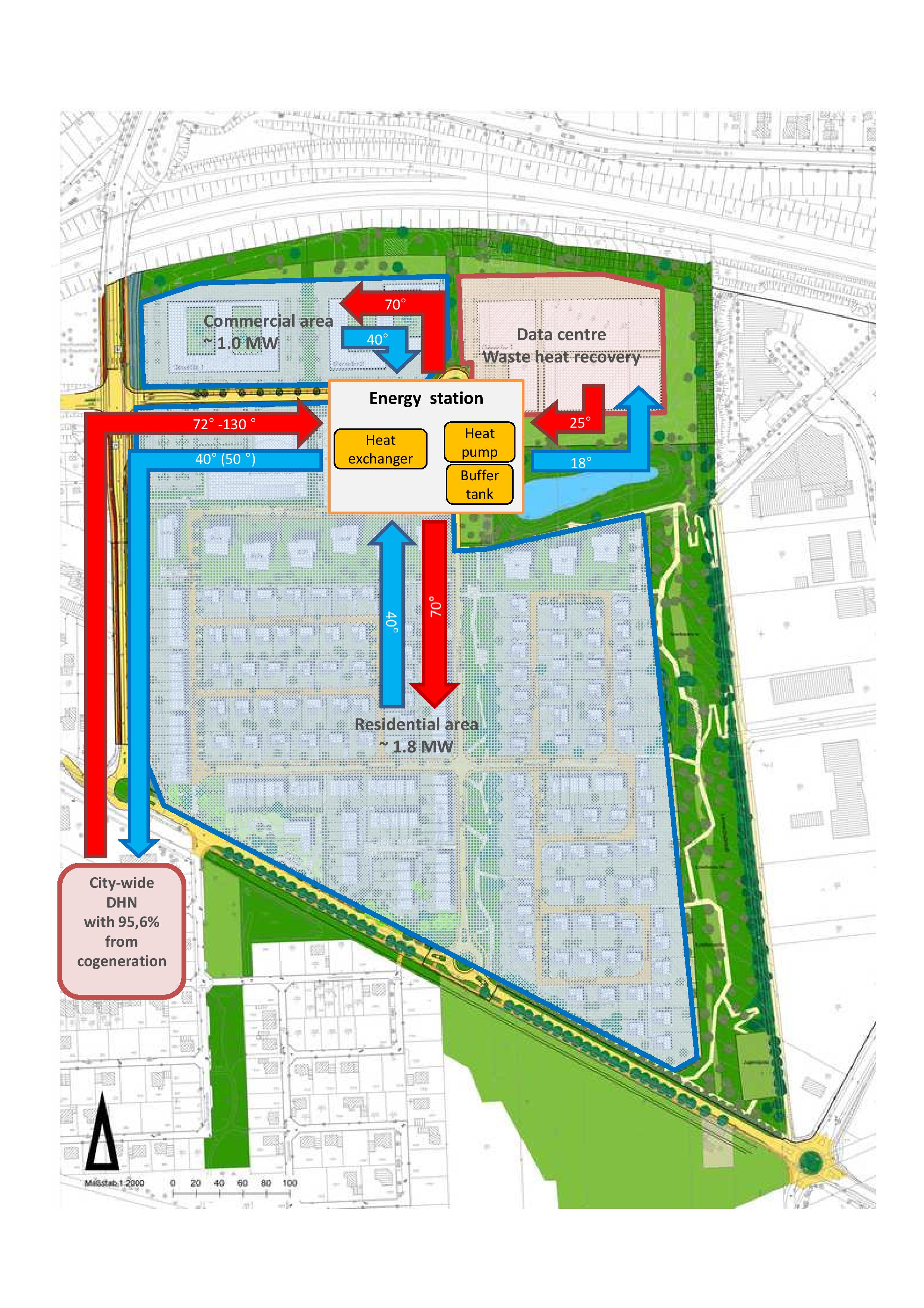}
\end{center}
\caption{An outline of the Brunswick demonstrator.}
\label{fig:Brunswick_sketch}
\end{figure}

We construct a simple model based on the Brunswick demonstrator in the aforementioned ReUseHeat project \cite{lygnerud2019contracts}.  The aim of the demonstrator is to supply heat to 400 newly built housing units, using waste heat from a nearby data centre which is upgraded to a suitable temperature with a heat pump.  The new housing units will also be connected to the existing city-wide network which, at present, supplies 45 percent of residents in the city using a gas powered combined heat and power (CHP) unit.  The intention is then that heat from the data centre will cover the baseline demand (demand that is present throughout the year such as for hot water) and heat from the CHP will cover seasonal demand.  Figure~\ref{fig:Brunswick_sketch} presents the layout of the Brunswick demonstrator.

For the study described in this paper, we plan to assess three different design options to meet the demand for heat. In the first, the entire heat demand is met by the CHP. The second follows the setup of the demonstrator for which baseline demand is met by heat from the data centre and seasonal demand is met by the CHP. In the third, the entire heat demand is met by waste heat from the data centre. 

To construct a model and run simulations, we employ an open source optimisation model for energy planning called Open Source Energy Modeling System (OSeMOSYS) \citep{Howells2011}. OSeMOSYS is a deterministic, linear optimisation model that obtains the energy supply mix that minimises the Net Present Cost (NPC), subject to a number of chosen constraints. 
We have chosen OSeMOSYS for two main reasons. Firstly, it provides us with the flexibility to operate on a local (city) level and to define our own scenarios. Secondly, OSeMOSYS is an open source model and therefore freely available for comparative project modelling.

\FloatBarrier
\FloatBarrier

\section{Uncertainty and Stochastic orderings}
\label{sec:StochasticOrdering}
We consider the use of stochastic ordering for describing and comparing uncertainty. The rationale is that stochastic orderings are weaker than a limited list of specific metrics. In particular, stochastic ordering is a partial ordering, whereas the commonly used metrics of uncertainty, such as standard deviation, impose total ordering.

\begin{defn}{\cite{Marshall1979}} \label{defn:stochastic_dom} A random variable $X$ is said to be stochastically less than (or equal to) $Y$, written $X\prec_{st} Y$, if the upper tail probabilities satisfy
\begin{equation}
P(X>t)\leq P(Y>t), \quad t\in\mathbb{R},    
\end{equation}
or
\begin{equation}
    P(X\leq t)\geq P(Y\leq t), \quad t\in\mathbb{R}.
\end{equation}
\end{defn}
We refer to this ordering as first-order stochastic dominance (SD). Marshall and Arnold \cite{Marshall1979} presented the following conditions equivalent to $X\prec_{st}Y$:
\begin{enumerate}[(i)]
  \item ${\mathbb E}\big[ g(X)\big] \leq {\mathbb E}\big[ g(Y)\big]$ for all increasing (non-decreasing) functions $g(\cdot)$;
  \item $g(X)\prec_{st}g(Y)$ for all increasing (non-decreasing) functions $g(\cdot)$;
  \item $P(X\in A)\leq P(Y\in A)$ for all sets $A$ with increasing indicator functions.
\end{enumerate}
The class of functions $g(\cdot)$ that satisfies (i.) and (ii.) is said to be the set of order-preserving functions with respect to a stochastic ordering.

From Definition \ref{defn:stochastic_dom}, we conclude that, to look for first-order SD, we can simply plot the empirical cdfs. If one always lies below the other, we can claim that it dominates with respect to any increasing function.

Consider NPC or emissions levels associated with two energy system designs, which we will refer to as design options $A$ and $B$. We construct a probability distribution for each output under each design option. Design option $A$ has first-order stochastic SD over design option $B$ if:
\begin{align*}
&F_A(x)\leq F_B(x)\quad\text{for all }x, \\
&F_A(x)<F_B(x)\quad\text{for at least one value of }x,
\end{align*}
where $F_A(x)$ and $F_B(x)$ are cdfs of NPC or emissions levels associated with design options $A$ and $B$, respectively. Since dominance in this example is associated with higher values of NPC or emissions levels, and our objective is to minimise the costs and the level of emissions, we are inclined to choose design option $B$ over design option $A$.\par

\subsection{Dispersion orderings}
We consider dispersion orderings (sometimes known as dispersive orderings), which are a particular case of stochastic orderings. Dispersion measures the extent to which a distribution is stretched or squeezed, and therefore dispersion orderings compare the spread of the probability distributions \cite{Shaked1982, Ayala2015}. Dispersion orderings can be extended for the multivariate variable case based on the idea of independent copies.

Let $\bb{X}$ and $\bb{Y}$ be two $d$-dimensional random vectors, i.e. $\bb{X}=(X_1, \dots, X_d)$ and $\bb{Y}=(Y_1, \dots, Y_d)$, with (multivariate) cdfs $F_{\bb{X}}(\bb{x})$ and $F_{\bb{Y}}(\bb{y})$. Let $D(\bb{r}, \bb{s})$ be some distance function between two $d$-dimensional vectors $\bb{r}$ and $\bb{s}$. Then, as above, let $(\bb{X}, \bb{X}')$ and $(\bb{Y}, \bb{Y}')$ be pairs of independent random vectors from the distributions $F_{\bb{X}}(\bb{x})$ and $F_{\bb{Y}}(\bb{y})$, respectively. We define a dispersion ordering $\prec_D$ to be $\bb{X}\prec_D\bb{Y}$ if and only if
\begin{equation}
    D(\bb{X}, \bb{X}')\prec_{st} D(\bb{Y}, \bb{Y}'),
\end{equation}
which says that $\bb{X}$ is less dispersive than $\bb{Y}$. In other words, less uncertainty is associated with the distribution of the vector $\bb{X}$ than the vector $\bb{Y}$.

Prior to describing particular dispersion orderings, we note that, for any given distance $D(\bb{r}, \bb{s})$, the class of order-preserving functions comprises all increasing (non-decreasing) functions, $g(\cdot)$ of $D(\bb{r}, \bb{s})$. By choice of distance and function, $g(\cdot)$, we can therefore cover a large range of dispersion metrics. \par

Consider the $L_2$ ordering introduced by \cite{giovagnoli1995multivariate} defined as 
$$D(\bb{r},\bb{s}) = ||\bb{r}-\bb{s}|| = \left\{\sum_{i=1}^d (r_i- s_i)^2 \right\}^{\frac{1}{2}},$$ 
where $D(\bb{r}, \bb{s})$ is the Euclidean distance, or $L_{2}$.  This is called the ``weak dispersion ordering''. Another dispersion ordering is the $L_1$ distance ordering determined from the following expressions:
$$ D(\bb{r},\bb{s}) = \sum_{i=1}^d |r_i- s_i|, $$
where $D(\bb{r}, \bb{s})$ is the $L_{1}$ distance.  

A natural extension of the independent copies idea is to take $k+1$ copies $\bb{X}_1, \ldots, \bb{X}_{k+1}$ and define a function, $\phi(\bb{X}_1,\bb{X}_2, \ldots, \bb{X}_{k+1})$, that describes the separation between them. In \cite{pronzato2017extended}, it is shown
that, if we take
$$\phi(\bb{x}_1, \ldots, \bb{x}_{k+1}) = (\mbox{vol} \{\triangle (\bb{x}_1, \ldots, \bb{x}_{k+1}) \})^2,$$ 
where $\triangle(\bb{x}_1, \ldots, \bb{x}_{k+1})$ is the $k$-dimensional simplex (in $d$ dimensions) whose vertices are $\bb{x}_1, \ldots, \bb{x}_{k+1}\in\mathbb{R}^d$, then  ${\mathbb E }\{ \phi(\bb{X}_1, \ldots, \bb{X}_{k+1})\}$
is a function of the variance-covariance matrix $\Sigma$ of the underlying distribution. This prompts a dispersion ordering $\bb{X} \prec_{\triangle} \bb{Y}$ defined by
$$ \mbox{vol} \{\triangle (\bb{X}_1, \ldots, \bb{X}_{k+1}) \} \prec_{st} \mbox{vol} \{\triangle (\bb{Y}_1, \ldots, \bb{Y}_{k+1}) \},$$
where $\bb{X}_1, \ldots, \bb{X}_{k+1}$ and $\bb{Y}_1, \ldots, \bb{Y}_{k+1}$ are independent draws from the $F_{\bb{X}}(\bb{x})$ and $F_{\bb{Y}}(\bb{y})$, respectively. The case in which $k=d$ was introduced by Oja \cite{oja1983descriptive} and discussed in \cite{giovagnoli1995multivariate} and is referred to as the Simplex ordering. Here, we remove the requirement that $k=d$ and refer to the approach as the Generalised Simplex ordering.  Note, again, that whenever we
see the ordering $\prec_{st}$, we can write down the class of order-preserving functions, $g(\cdot)$, in the present case of $\mbox{vol} \{\triangle (\bb{X}_1, \ldots, \bb{X}_{k+1}) \}$. Dispersion orderings based on Hausdorff distance \cite{lopez2006indexed} and Mahalanobis distance \cite{pronzato2018simplicial} are possible alternatives to the simplex ordering. \par

We briefly point out the impact of scaling on dispersion orderings. When considering the $L_1$ distance ordering, the scaling of the variables impacts the relative contribution of each one in the calculation of the $L_1$ distance. Therefore, the $L_1$ ordering is sensitive to scale and pre-processing of the data is required to ensure that we operate in the same range across all dimensions. On the contrary, the simplex volume is a scale-free, homogeneous measure and therefore no pre-processing is required for the Generalised Simplex ordering. For more details see \ref{subsec:scaling}. We also note that computing the simplex volume can become computationally expensive, in particular when we are operating with large data sets. To deal with this, \cite{Ayala2015} proposed the use of bootstrap resamples from the data set to obtain the distances. Despite this, we believe the properties of the Generalised Simplex ordering make it an attractive choice and we provide a demonstration of its use in Section \ref{sec:Results}, 

\subsection{Quantifying the difference between cdfs} \label{subsection:quantifying} 
It is useful to quantify the extent of the difference between two cdfs when one dominates another. Here, we use a Kolmogorov-Smirnov (KS) distance measure, which is defined as follows. Let the random variables $X$ and $Y$ have cdfs $F_{X} (z)$ and $F_{Y}(z)$ respectively.  The KS distance between $X$ and $Y$ is given by
$$
D=\sup_z|F_X(z)-F_Y(z)|, \quad z\in \mathbb{R}.
$$
Noting that the value of $D$ lies in the range $[0, 1]$, the statistic tells us the maximum difference between the two cdfs.  If $D$ is small, the extent to which one dominates the other is small and vice versa.

\subsection{Hypothesis testing for stochastic orderings} \label{section:hypothesis_test}

In the studies presented in Section \ref{sec:Results}, the distributions of the outputs are produced using an energy systems simulation (OSeMOSYS \cite{Howells2011}) and therefore we can only approximate the underlying cdf of the output variables that would be produced with an infinite sample size. Thus we make use of the Kolmogorov-Smirnov (KS) test proposed by \cite{Mcfadden1989} which defines the null hypothesis to be the case that $Y$ first-order stochastically dominates $X$ and the alternative hypothesis that the null hypothesis is false. If the null hypothesis can be rejected, we do not have enough evidence that one dominates the other.

Formally, let $X_1, \dots, X_n$ and $Y_1, \dots, Y_m$ be samples from $F_X$ and $F_Y$, respectively. Let the null hypothesis be
$$
H_0: F_X(z)\geq F_Y(z), \quad z\in \mathbb{R}
$$
The alternative hypothesis is
$$
H_1: F_X(z)<F_Y(z), \quad z\in\mathbb{R}.
$$
{We perform a one-tailed Kolmogorov-Smirnov test and consider $D^{-}$, the maximum negative deviation, defined as
$$
D^{-}= \sup_z (F_Y(z)-F_X(z)).
$$
Large values of $D^{-}$ will occur when the alternative hypothesis is true and the null hypothesis can be rejected, providing significant evidence against stochastic ordering.  This approach is similar to that used by \cite{Ayala2015} who tested the Hausdoff and Simplex dispersion orderings.

We note that an assumption of the KS test is that the samples should be independent of each other and, although, in each sample, the simulations are performed independently, the input parameters induced by the scenarios are common to both and therefore the assumption fails. Similar technical difficulties arise with the bootstrap methodology.  Nonetheless, we note that the effect of this failed assumption is small and diminishes with sample size (which is generally large).  We therefore believe that these results are informative for our analysis and provide a useful diagnostic technique in general.

Another issue requiring us to be cautious is the use of multiple testing. In our experiments, we consider three options $(N=3)$ and, as a result, we perform three pairwise post-hoc tests to assess the differences between them.  To account for this multiple testing, we make use of the Bonferroni correction.  Therefore, if we specify a significance level of $\alpha=0.05$, the corresponding significance level for each individual hypothesis is adjusted to $\alpha/N=0.05/3=0.0167$.




\section{Experimental design}
\label{sec:Experiments}
We demonstrate the use of stochastic ordering to assess the aforementioned waste heat recovery project in Brunswick (see Section~\ref{section:brunswick} for more details).  Three different design options are considered, each of which differ according to the technology mix employed for supplying domestic heat.  These are outlined in Table~\ref{tab:Design}.  Each design option is evaluated in terms of its Net Present Cost (NPC) (mln \euro{}) and $CO_2$-equivalent emissions (in metric tonnes). Here, we analyse the uncertainty in the outputs of the model induced by variations in four different inputs:
\begin{enumerate}
\item Operational costs.
\item Discount rate.
\item Coefficient of Performance (COP) for the heat pump (units of heat delivered per unit of electricity).
\item Emission Activity Factor (the emissions produced (in metric tonnes) from operating a particular technology in the energy system).
\end{enumerate}
Variations in the four input variables are expected to impact both the NPC and emissions levels.  For each input variable, we specify three levels: low, medium and high. We then perform simulations with a full factorial design (often known as a fully crossed design) so that all possible combinations across the model inputs are considered \citep{George2005}. This gives a total of 81 simulation runs. \par

\FloatBarrier
\begin{table}[ht]
\centering 
\begin{tabular}{| m{4cm} | m{9cm} | } 
\hline 
\textbf{Design type} & \textbf{Description}\\ [0.5ex]
\hline
Design Option 1 & Combined Heat and Power (CHP) is employed to meet both the baseload and seasonal heat demand. \\
\hline
Design Option 2 & A heat pump is employed to meet baseload heat demand and CHP is used to meet seasonal heat demand. \\
\hline
Design Option 3 & A heat pump is employed with a small amount of storage to meet both the baseload and seasonal heat demand. \\
\hline
\end{tabular}
\caption{Description of design options in the study.} 
\label{tab:Design}
\end{table}
\FloatBarrier
We consider each of the design options under three different scenarios.  Scenario Analysis is a widely used uncertainty tool in energy systems studies \cite{Wheatcroft2019}. Examples of scenarios related to energy include the Future Energy Scenarios (FES) published by the UK National Grid \cite{FES2021} and World Energy Scenarios \cite{WorldEnergy2019}, both of which consider different pathways to decarbonisation.

We define three scenarios that differ in terms of selected elements of government climate policy and consumer engagement with green technology. A description of the three scenarios is given in Table~\ref{tab:Scenarios}. In the Green scenario, a combination of high consumer engagement and government incentive schemes are used to rapidly reach net zero. In the Market scenario, there is a reliance on market forces and only limited government intervention in the energy system planning. Finally, the Neutral scenario captures the middle ground between the two. Since the operational lifetime of a typical heat pump is around twenty years, this time horizon is used for the study.

Under each scenario, we produce model simulations that, for a given set of model inputs, generate the volume of emissions and NPC over a 20 year period as model outputs.  In Section \ref{subsec:order_mean}, we consider the empirical cdfs of the model outputs under different scenarios and design options. Then, in Section \ref{subsec:dispersion_order}, we consider the Generalised Simplex dispersion ordering to compare the effects of different scenarios and design options on uncertainty.

\FloatBarrier
\begin{table}[ht]
\centering 
\begin{tabular}{| m{4cm} | m{9cm} | } 
\hline 
\textbf{Scenario} & \textbf{Description} \\ 
\hline
Green  scenario & Penalty per Mton of emissions: 100 \euro{} per Mton.
 
Annual change in baseload and seasonal heat demand: -1\%.

Increasing gas prices and decreasing electricity prices.\\
\hline
Neutral  scenario & 
Penalty per Mton of emissions: 40 \euro{} per Mton.

Baseload and seasonal heat demand fluctuate around the central projections.

Gas and electricity prices stay within the central projected values.\\
\hline
Market scenario & 
No penalty per Mton of emissions.

Annual change in baseload and seasonal heat demand: 1\%.

Decreasing gas prices and increasing electricity prices.\\
\hline
\end{tabular}
\caption{Description of the selected scenarios in the study.} 
\label{tab:Scenarios}
\end{table}
\FloatBarrier

\section{Results}
\label{sec:Results}
Figure \ref{fig:ScatterPlot} shows scatter plots of the two model outputs obtained for each of the three design options and under each scenario. Focusing first on emissions, since the emissions factor for natural gas is higher than for electricity, the highest level of emissions is produced by design option 1, followed by design option 2 and then design option 3. From the second row in Figure~\ref{fig:ScatterPlot}, we observe that the different scenarios have only a small impact on the levels of emissions under each of the design options. This is not surprising since government interventions typically aim at changing behaviour through cost. \par

Differences in the three scenarios have a much larger impact on NPC. Under the Market and Neutral scenarios, design option 1 corresponds to the lowest NPC values, which can be explained by the high investment cost of the heat pump. However, under the Green scenario, the opposite is true due to a high carbon penalty. Therefore, under the Green scenario, design options 2 and 3 become more attractive alternatives, thus demonstrating the value of considering different scenarios when informing decision-makers. 

\FloatBarrier
\begin{figure}[ht]
\begin{center}
\includegraphics[width=0.8\textwidth,height=0.3\textheight]{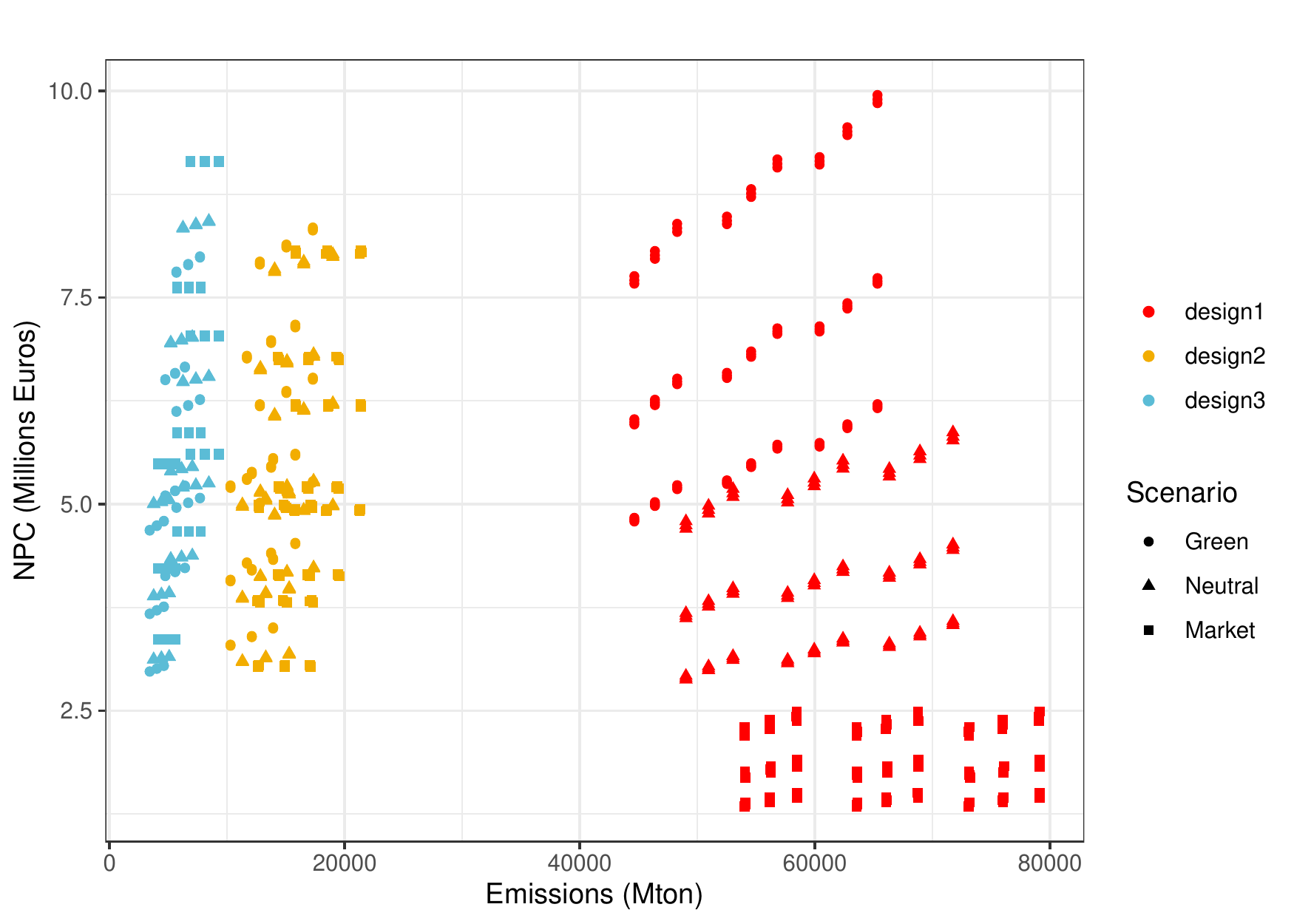}
\includegraphics[width=0.8\textwidth,height=0.3\textheight]{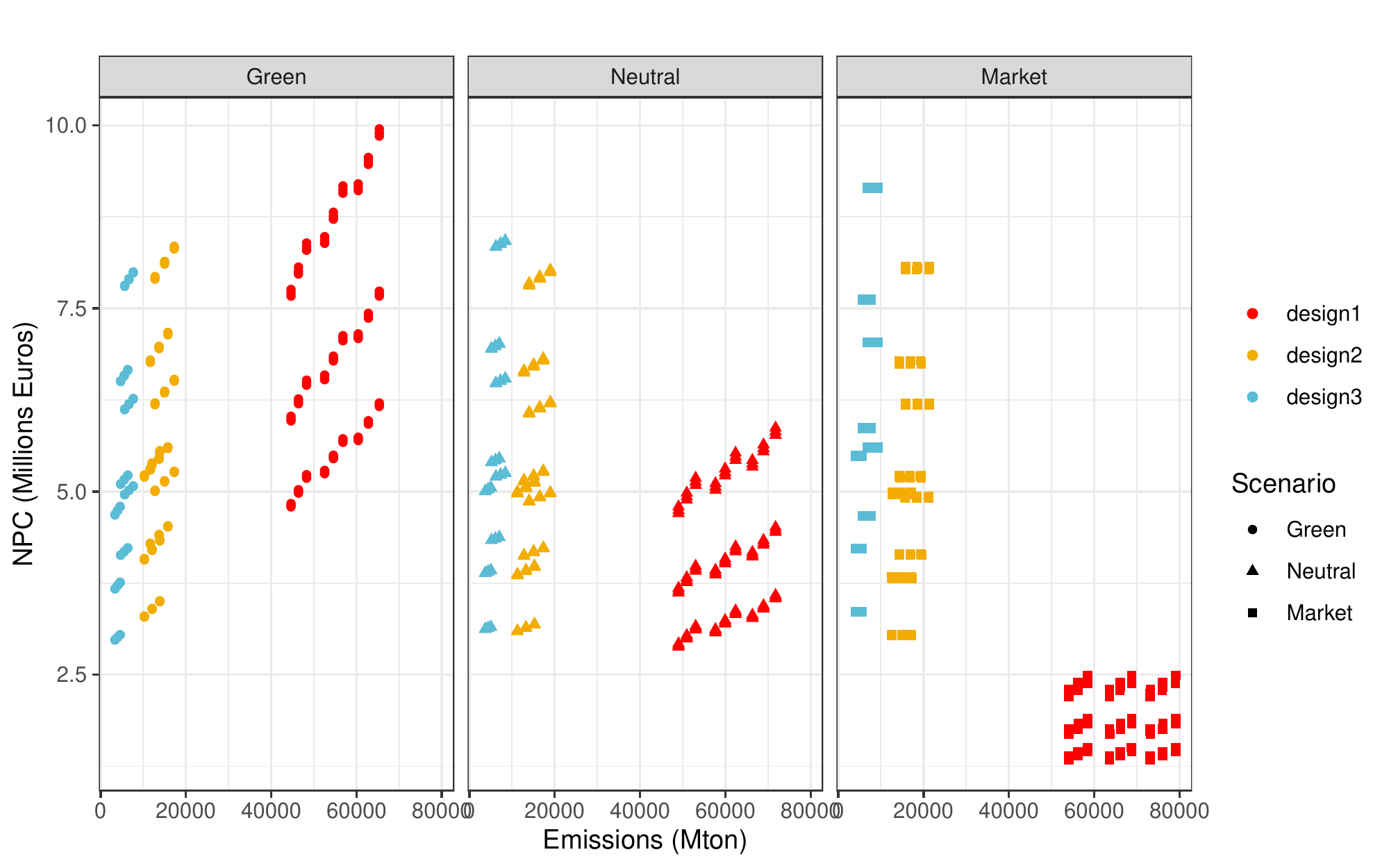}
\end{center}
\caption{\textit{First row}: NPC (mln \euro) against $CO_2$-equivalent emissions (Mton) for all three design options under the three scenarios. \textit{Second row}: NPC against emissions for all three design options, plotted separately for each scenario.}
\label{fig:ScatterPlot}
\end{figure}
\FloatBarrier

\subsection{Orderings in the mean}
\label{subsec:order_mean}
Figure \ref{fig:Plot1NPC} shows the cdfs of the NPC of each design option under each of the three scenarios (first row) and of each scenario under the three design options (second row).  In Table \ref{tab:KS_first_order_design} values of the Kolmogorov-Smirnov (KS) distance measure between the cdfs of different design options under each scenario are shown.  The numbers in the table represent the distance between the cdfs when the design option denoted in the column dominates that denoted in the row. In each cell, there are three numbers that correspond to the Market, Neutral and Green scenarios, respectively. Cases in which there is no dominance in this direction are denoted with a dash.  In addition, we perform the one-tailed Kolmogorov-Smirnov test defined in Section \ref{section:hypothesis_test} to assess whether the null hypothesis of stochastic dominance can be rejected with $p$-values corresponding to NPC and emissions shown in Table \ref{tab:KS_first_order_design_pvals}. Cases where the cdfs cross and first-order SD clearly does not exist are denoted by an `na'.

From the first row of Figure \ref{fig:Plot1NPC}, we observe that, under the Market scenario, design option 3 has first-order SD over design option 2 which has first-order SD over design option 1. Under the Green scenario, the ordering is reversed, mainly due to the increased financial support from policy makers for renewable energy sources. Under the Neutral scenario, the cdfs for design option 2 and design option 3 cross, so first-order SD cannot discriminate between these two options. However, we observe that design option 1 is dominated by the other two options. We therefore conclude that, if the sole aim is to minimise NPC, under the Green scenario, design option 3 would be the preferred option whilst design option 1 would be the preferred option under the Neutral and Market scenarios.  The reason for this difference is that, under the Green scenario, the government introduces high penalties for generation of emissions, which makes design option 1 an unattractive option for providing heat.  

The KS values in Table \ref{tab:KS_first_order_design} demonstrate the extent to which there is dominance between design options.  For example, in the Market and Neutral scenarios, design option 2 dominates design option 1 with a KS distance of $1$ and $0.37$, respectively. In the Green Scenario, the dominance is reversed and design option 1 dominates design option 2 with a KS distance of $0.48$.  Focusing on Table \ref{tab:KS_first_order_design_pvals}, for NPC, the $p$-values for whether design option 1 dominates design option 2 are $0/<0.001/1$.  This means that stochastic dominance in this direction can be rejected for the Market and Neutral scenarios whilst it is not rejected for the Green Scenario. 



\begin{table}[ht]
\begin{center}
\begin{tabular}{ |c|c|c|c|c|c| } 
\hline
 & & design 1 & design 2 & design 3\\
\hline
\multirow{3}{4em}{NPC} & design 1 & & 1/0.37/- & 1/0.41/-\\ 
& design 2 & -/ -/0.48 & &  0.33/NA/-\\
& design 3 & -/-/0.57 & -/NA/0.23 & \\ 
\hline
\multirow{3}{4em}{Emissions} & design 1 & & -/-/- & -/-/-\\ 
& design 2 & 1/1/1 & &  -/-/-\\
& design 3 & 1/1/1 & 1/1/1 & \\ 
\hline
\end{tabular}
\caption{Kolmogorov-Smirnov distances between empirical cdfs of NPC and $CO_2$-equivalent emissions for pairs of design options. The KS values correspond to the Market, Neutral and Green scenarios and are only shown when the cdf for the design option in the row is dominated by that in the column. NA values indicate that the cdfs cross, and first-order SD does not exist in either direction.}
\label{tab:KS_first_order_design}
\end{center}
\end{table}

\begin{table}[ht]
\begin{center}
\begin{tabular}{ |c|c|c|c|c|c| } 
\hline
 & & design 1 & design 2 & design 3\\
\hline
\multirow{3}{4em}{NPC} & design 1 & & 1/1/$0^*$ & 1/1/$0^*$\\ 
& design 2 & $0^*$/ $<0.001^*$/1 & &  1/NA/$0.012^*$\\
& design 3 & $0^*$/$<0.001^*$/$<0.001^*$ & $<0.001^*$/NA/1 & \\ 
\hline
\multirow{3}{4em}{Emissions} & design 1 & & $0^*$/$0^*$/$0^*$ & $0^*$/$0^*$/$0^*$\\ 
& design 2 & 1/1/1 & &  $0^*$/$0^*$/$0^*$\\
& design 3 & 1/1/1 & 1/1/1 & \\ 
\hline
\end{tabular}
\caption{Observed $p$-values for NPC and $CO_2$-equivalent emissions using first-order stochastic ordering. The three $p$-values in each table entry correspond to the Market, Neutral and Green scenarios and an asterisk denotes that stochastic dominance is rejected at the chosen significance level. NA values indicate that the cdfs cross, and  therefore no hypothesis test is performed.}
\label{tab:KS_first_order_design_pvals}
\end{center}
\end{table}

Focusing now on the effects of each scenario on the three design options, under design option 1 the Green scenario dominates the Neutral scenario which dominates the Market scenario. In this case, the ordering indicates that the highest NPC values are generated under the Green scenario.  For design option 3, on the other hand, the ordering between scenarios is reversed.  Under design option 2, the cdf for the Neutral scenario dominates that of the Green scenario whilst the cdf for the Market scenario crosses the other two cdfs and thus there is no ordering.  Overall, the difference in orderings may lead a planner to think carefully about their choice of design option under different possible futures.  If, for example, they consider policy featured in the Green scenario to be likely, they may choose design option 2 or 3 to mitigate that risk. 

In Table \ref{tab:KS_first_order_scenario}, we present KS distances for the NPC and emissions associated with each scenario under the selected design options.  Table \ref{tab:KS_first_order_pvals_scenario} provides $p$-values for the one-sided KS test. Notably, for NPC, under design option 1, we observe high values of the KS distance between pairs of scenarios. This demonstrates a substantial scenario effect which is much larger than for the other two design options.  This is reflected in the high $p$-values which indicate that SD cannot be rejected.


Figure \ref{fig:Plot1Emissions} shows the empirical cdfs corresponding to emission levels. The empirical cdfs of emissions for each design option under each of the three scenarios (top row) demonstrate that, under all three scenarios, design option 1 dominates design option 2, which dominates design option 3.  This, of course, is unsurprising since the scenarios differ in terms of the extent to which policy is driven by `green' considerations and confirms our expectation that waste heat recovery is a carbon reducing technology under most reasonable policy decisions.

\begin{table}[ht]
\begin{center}
\begin{tabular}{ |c|c|c|c|c|c| } 
\hline
 & & Market & Neutral & Green\\
\hline
\multirow{3}{4em}{NPC} & Market & & 1/NA/- & 1/NA/-\\ 
& Neutral & -/ NA/0.33 & &  0.78/0.20/-\\
& Green & -/NA/0.33 & -/-/0.18 & \\ 
\hline
\multirow{3}{4em}{Emissions} & Market & & -/-/- & -/-/-\\ 
& Neutral & 0.33/0.33/0.22 & &  -/-/-\\
& Green & 0.56/0.56/0.44 & 0.33/0.33/0.22 & \\ 
\hline
\end{tabular}
\caption{Kolmogorov-Smirnov distances between empirical cdfs of NPC and $CO_2$-equivalent emissions for pairs of scenarios. The KS values correspond to design option 1, design option 2 and design option 3 and are only shown when the cdf for scenario in the column lies below that in the row. NA values indicate that cdfs cross, and first-order SD does not exist.}
\label{tab:KS_first_order_scenario}
\end{center}
\end{table}

\begin{table}[ht]
\begin{center}
\begin{tabular}{ |c|c|c|c|c|c| } 
\hline
 & & Market & Neutral & Green\\
\hline
\multirow{3}{4em}{NPC} & Market & & 1/NA/$<0.001^*$ & 1/NA/$<0.001^*$\\ 
& Neutral & $0^*$/NA/1 & &  1/1/0.06\\
& Green & $0^*$/NA/1 & $0^*$/0.042/1 & \\ 
\hline
\multirow{3}{4em}{Emissions} & Market & & $<0.001^*$/$<0.001^*$/0.018 & $0^*$/$0^*$/$0^*$\\ 
& Neutral & 1/1/1 & &  $<0.001^*$/$<0.001^*$/0.018\\
& Green & 1/1/1 & 1/1/1 & \\ 
\hline
\end{tabular}
\caption{Observed $p$-values for NPC and $CO_2$-equivalent emissions using first-order stochastic ordering. The $p$-values correspond to design option 1, design option 2 and design option 3, where $p$-values with asterisk are less than or equal to significance level. NA values indicate that cdfs cross, and first-order SD does not exist and we cannot perform hypothesis tests.}
\label{tab:KS_first_order_pvals_scenario}
\end{center}
\end{table}

\FloatBarrier
\begin{figure}[ht]
\begin{center}
\includegraphics[width=0.95\textwidth,height=0.3\textheight]{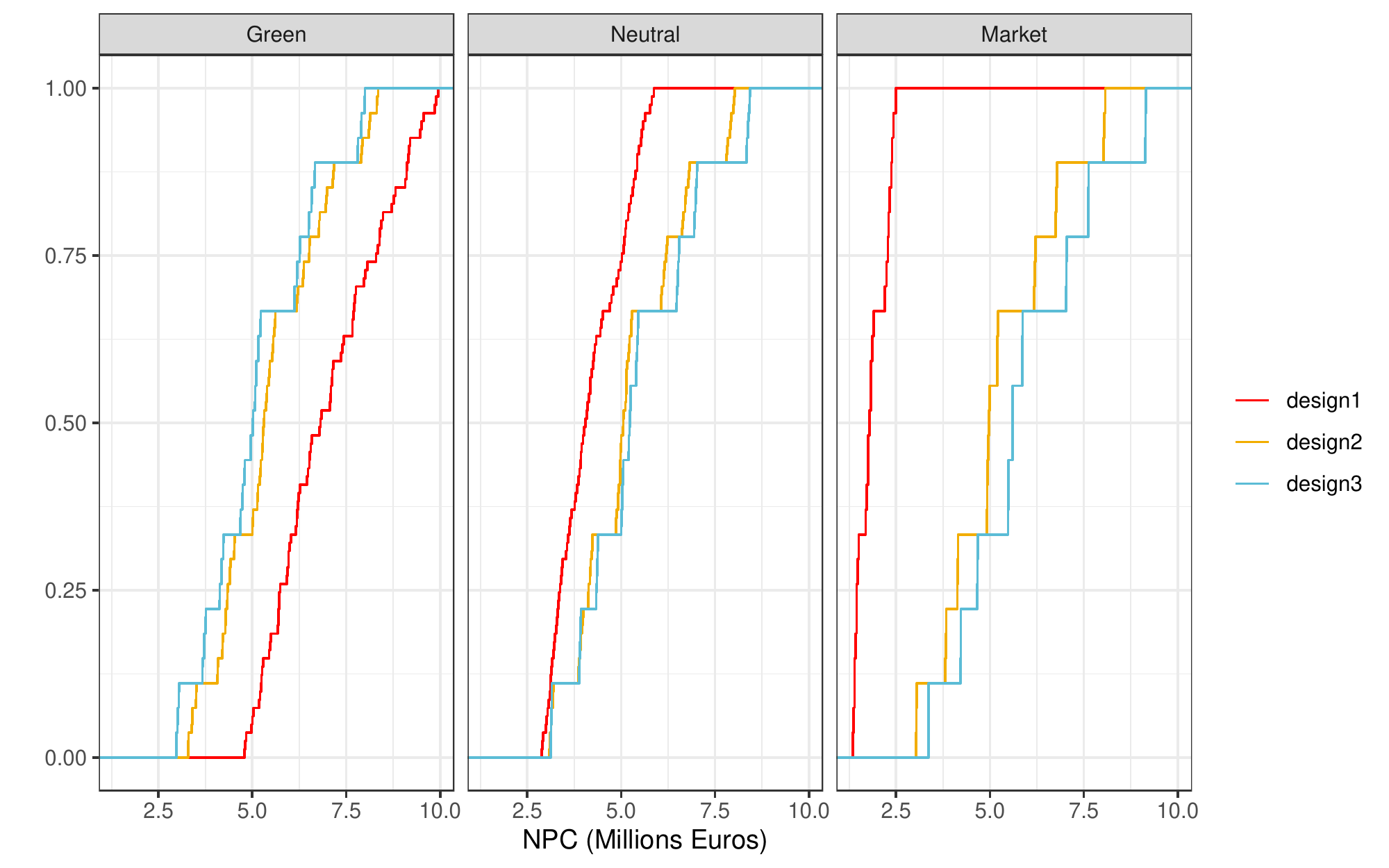}
\includegraphics[width=0.95\textwidth,height=0.3\textheight]{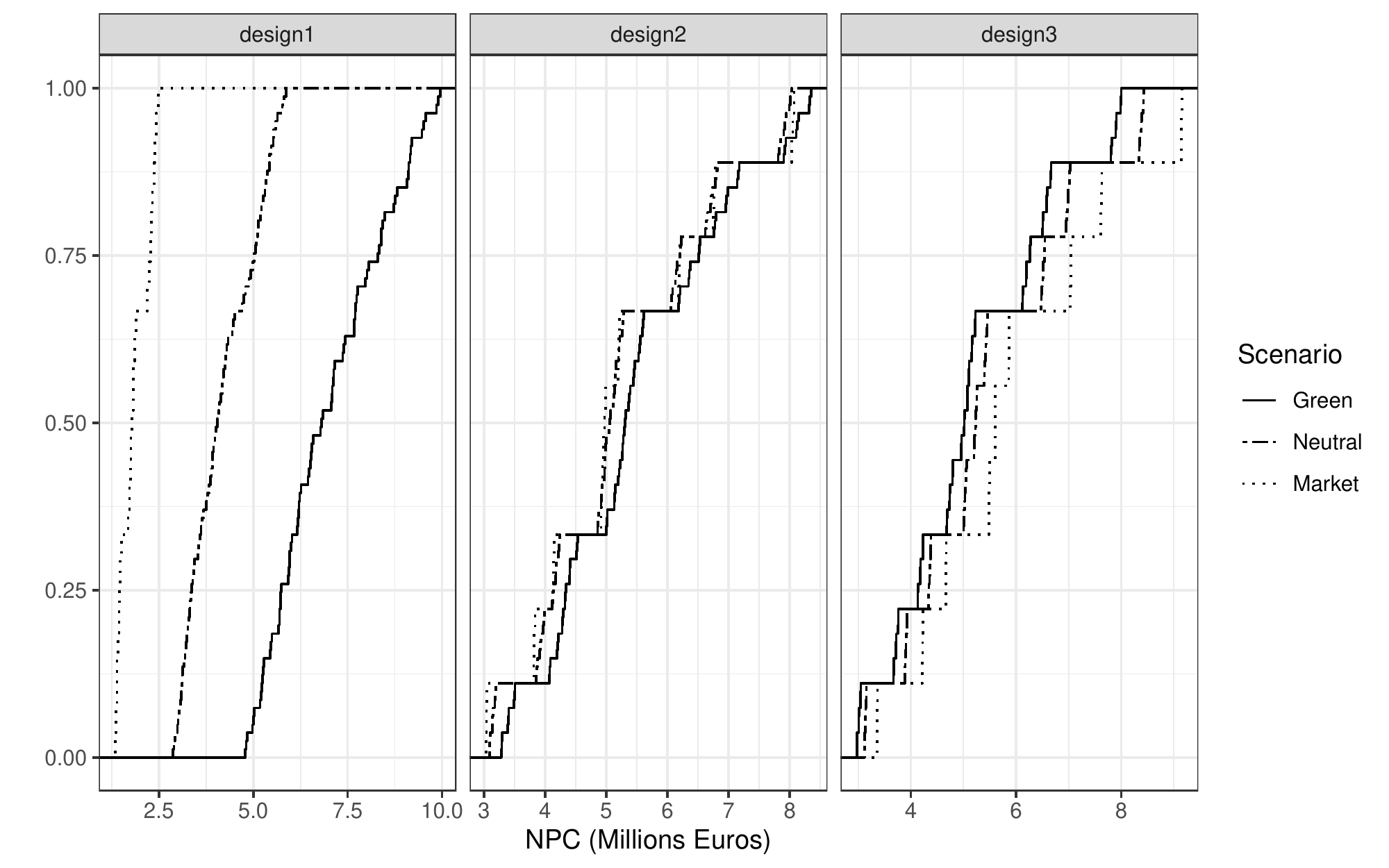}
\end{center}
\caption{Empirical cdfs for NPC for (i) all three design options plotted together for each individual scenario (\textit{first row}), (ii) all three scenarios plotted together for each individual design option (\textit{second row}).}
\label{fig:Plot1NPC}
\end{figure}
\FloatBarrier  

\FloatBarrier
\begin{figure}[ht]
\begin{center}
\includegraphics[width=0.9\textwidth,height=0.25\textheight]{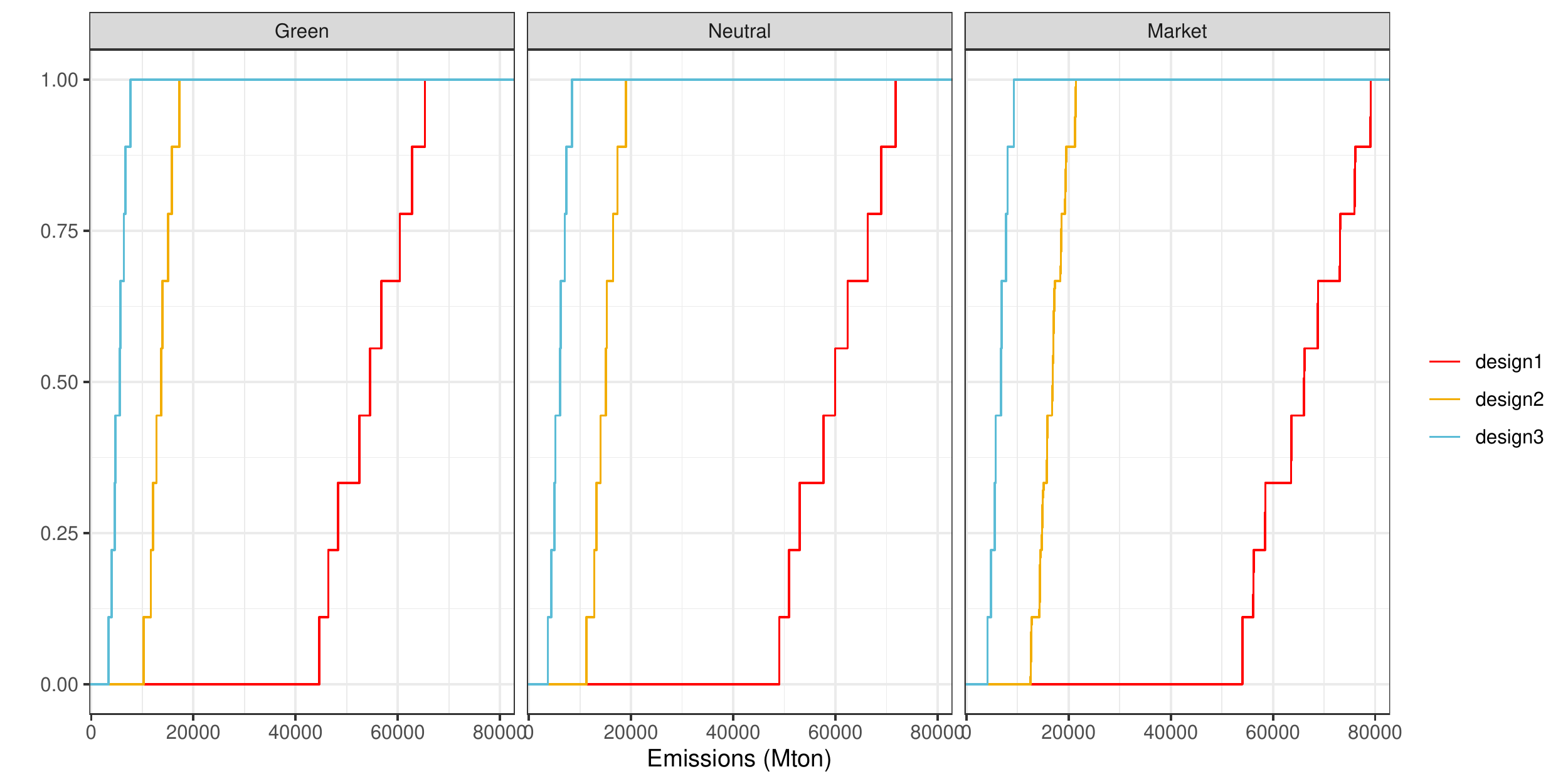}
\includegraphics[width=0.9\textwidth,height=0.25\textheight]{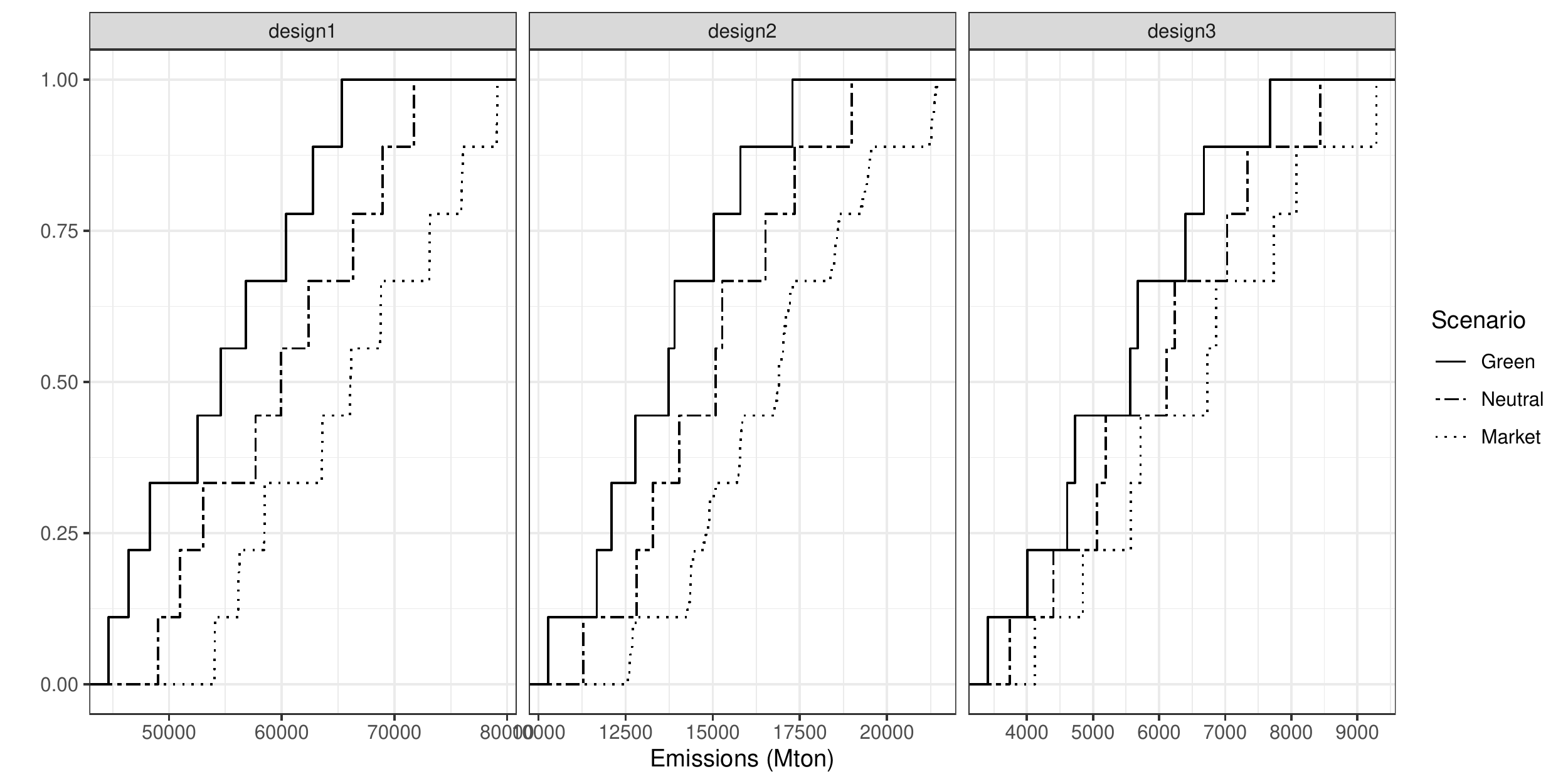}
\end{center}
\caption{Empirical cdfs for $CO_2$-equivalent emissions for (i) all three design options plotted together for each individual scenario (\textit{first row}), (ii) all three scenarios plotted together for each individual design option (\textit{second row}).}
\label{fig:Plot1Emissions}
\end{figure}
\FloatBarrier

\subsection{Generalised Simplex dispersion ordering}
\label{subsec:dispersion_order}
In Section \ref{subsec:order_mean}, we demonstrated how to assess the impacts of different design options by considering orderings in the mean. We now focus on dispersion, and note that a dispersive probability distribution indicates a high level of uncertainty (risk), which is assumed to be a negative attribute.  Here, we focus primarily on the effects of scenarios on each of the design options. 

Figure \ref{fig:Plot4} displays empirical cdfs for the Generalised Simplex ordering.  Those of the three design options under each of the three different scenarios are shown in the top row, whilst those of the three scenarios under each of the three design options are shown in the second row. Table \ref{tab:KS_GS_design} shows the KS distances between the empirical cdfs associated with the pairs of design options under the Market, Neutral and Green scenarios, respectively, and the $p$-values in Table \ref{tab:pvals_GS_design} correspond to the KS tests of the dispersion ordering. 

Under the Green and Neutral scenarios there is a clear ordering: design option 3 is lesser than design option 2 which is lesser than design option 1 in the Generalised Simplex dispersion order. Under the Market scenario, design option 3 is dominated by the other two design options, though the cdfs of design options 1 and 2 cross. Under the Generalised Simplex ordering, design option 3 therefore shown to be the most robust option under all three scenarios, whilst, in the Green and Neutral scenarios, design option 2 is more robust than design option 1. The KS distances between the design options in Table \ref{tab:KS_GS_design} are smallest under the Market Scenario and largest under the Green Scenario, suggesting that the difference in robustness is largest for the case in which government prioritise green issues. In all cases in which SD is demonstrated, the difference according to the KS test is significant.  

For a planner, the above results are informative in that they demonstrate the relative robustness of different design options under each of the three scenarios, which is an important criterion for decision-makers.  
\begin{table}[ht]
\begin{center}
\begin{tabular}{|c|c|c|c|c| }
\hline
& design 1 & design 2 & design 3\\
\hline
design 1 & & NA/-/- & -/-/-\\
design 2 & NA/0.33/0.49 & &  -/-/-\\
design 3 & 0.19/0.49/0.63 & 0.19/0.21/0.24 & \\
\hline
\end{tabular}
\caption{Kolmogorov-Smirnov distances between Generalised Simplex empirical cdfs for pairs of design options. The KS values correspond to the Market, Neutral and Green scenarios and are only shown when the cdf for design option in the column lies below that in the row. NA values indicate that cdfs cross, and first-order SD does not exist.}
\label{tab:KS_GS_design}
\end{center}
\end{table}

\begin{table}[ht]
\begin{center}
\begin{tabular}{|c|c|c|c|c|}
\hline
& design 1 & design 2 & design 3\\
\hline
design 1 & & NA/$0^*$/$0^*$ & $0^*$/$0^*$/$0^*$\\
design 2 & NA/0.98/1 & &  $0^*$/$0^*$/$0^*$\\
design 3 & 1/1/1 & 1/1/1 & \\
\hline
\end{tabular}
\caption{Observed $p$-values for different design options using the Generalised Simplex dispersion ordering. The $p$-values correspond to the Market, Neutral and Green scenarios, where those with an asterisk are less than or equal to the chosen significance level. NA values indicate that cdfs cross, and first-order SD does not exist and we cannot perform hypothesis tests.}
\label{tab:pvals_GS_design}
\end{center}
\end{table}

\FloatBarrier
\begin{figure}[ht]
\begin{center}
\includegraphics[width=1\textwidth,height=0.25\textheight]{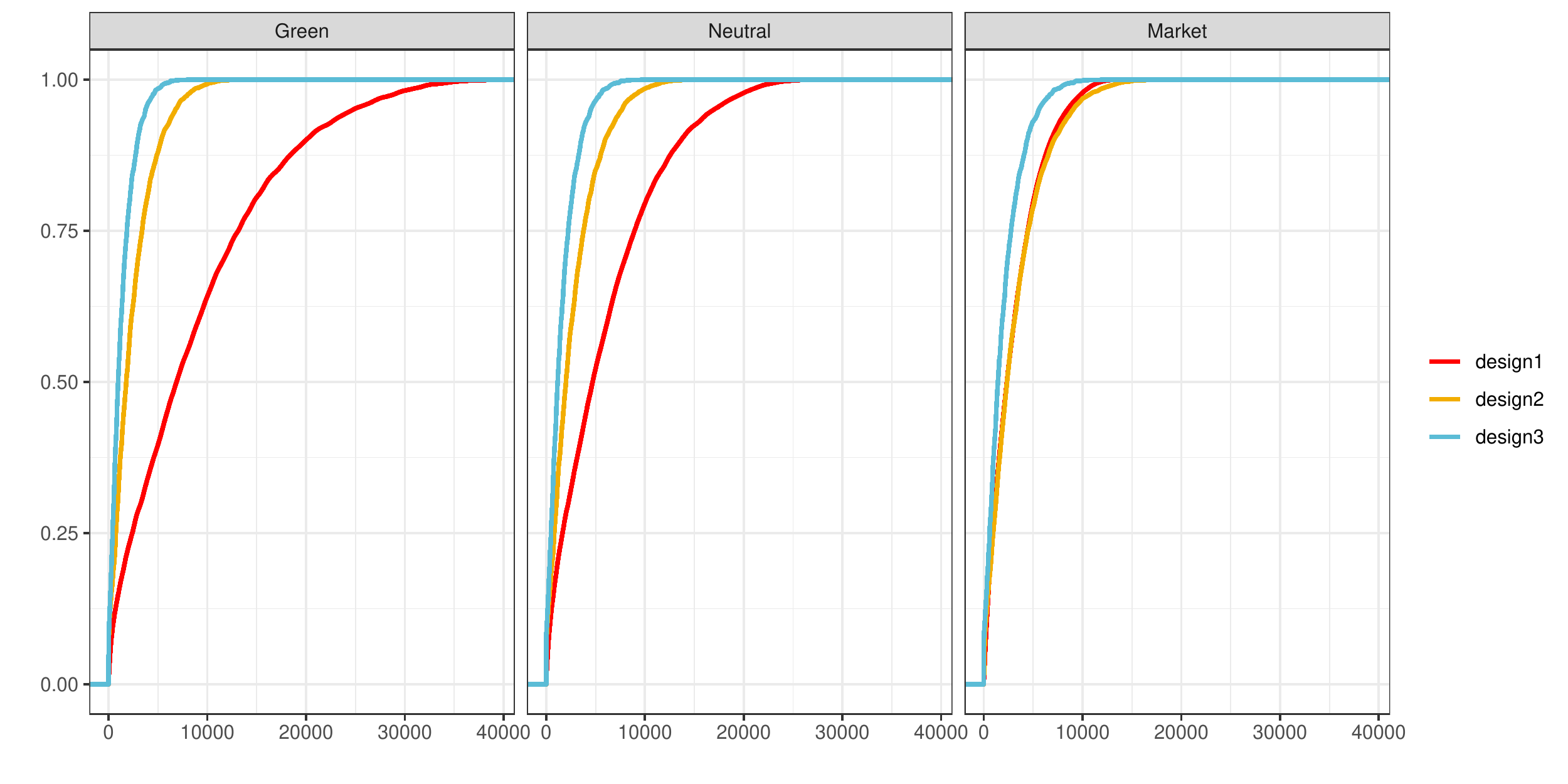}
\includegraphics[width=1\textwidth,height=0.25\textheight]{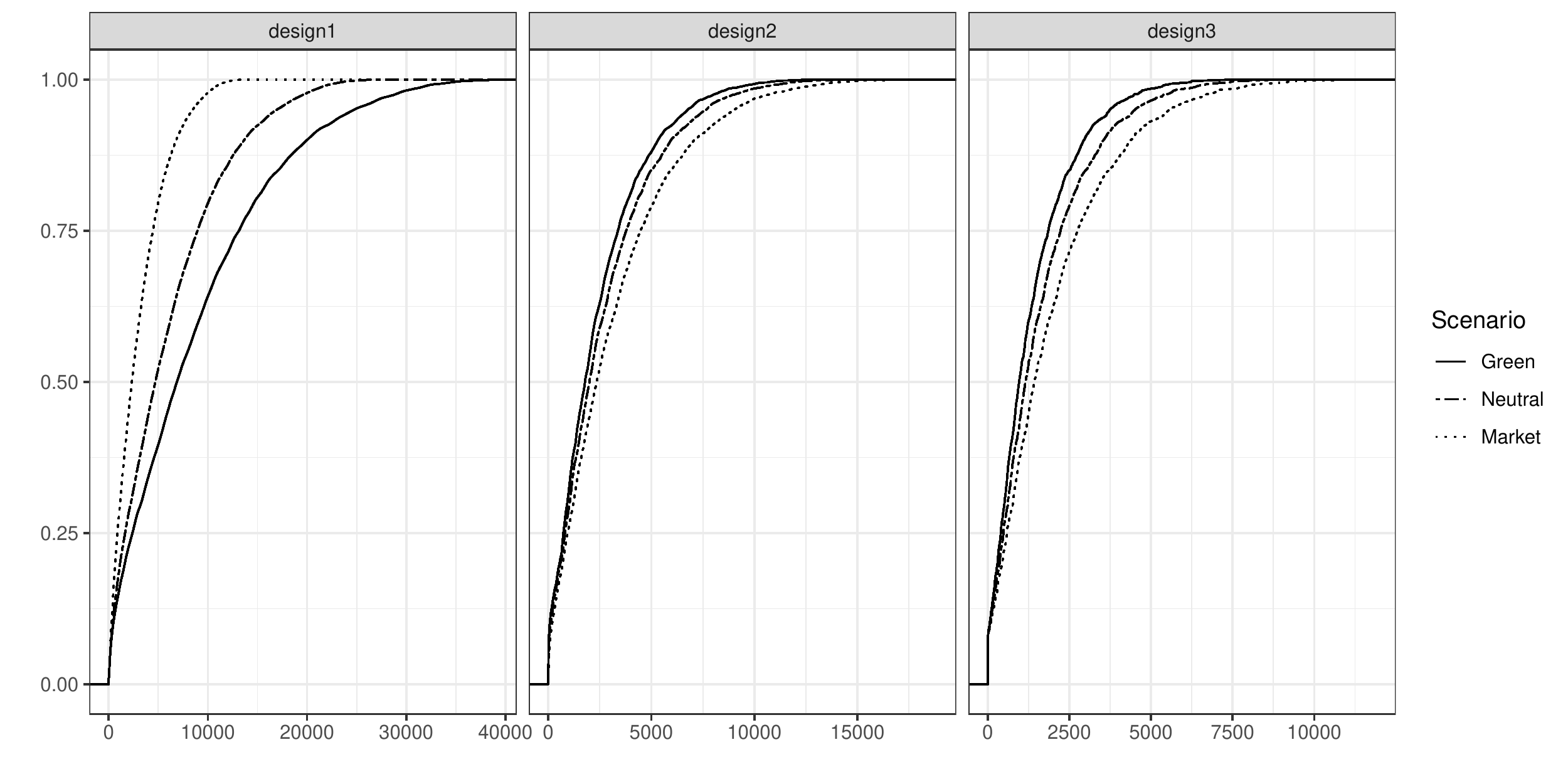}
\end{center}
\caption{\textit{First row}: empirical cdfs of the Generalised Simplex metric for all three design options plotted together for each individual scenario. \textit{Second row}: empirical cdfs of the Generalised Simplex metric for all three scenarios plotted together for each individual design option.}
\label{fig:Plot4}
\end{figure}
\FloatBarrier

\section{Discussion} \label{sec:Discussion}
The importance of informing decision-makers and stakeholders about the uncertainty associated with design choices is increasingly recognised in energy systems \cite{Soroudi2013}. The present paper demonstrates the use of stochastic ordering (dominance) in the context of local energy planning under uncertainty. In particular, we have introduced variability at two levels: local variability in inputs and more general scenarios to account for changes in consumer behaviour and the political environment. Based on a generated data set, we have considered stochastic orderings both in shift and dispersion.
 
In our analysis of the waste heat recovery system in Brunswick, we have demonstrated that, under a Green scenario, which assumes active government policy to meet the 2050 net-zero carbon target, lower $CO_2$-equivalent emissions can be produced at a lower cost by employing a heat pump with heat from a data centre. In addition, we have found using the Generalised Simplex ordering that this design choice is more robust (less volatile) compared to the other options under all three scenarios.  We argue that, if it can be shown to be true in general, this robustness is an attractive feature of low-temperature waste heat recovery. We note that, in a number of cases, the cdfs cross and therefore first-order SD failed to discriminate between these respective choices. One possible next step is to consider higher-orders of stochastic dominance, which can be performed as an extension to the presented analysis.
 
We consider the proposed approach for analysing and comparing different options for providing residential heat in district heating as a competitive alternative to traditional uncertainty metrics used in energy systems.  Our non-parametric method offers orderings in uncertainty based on distribution functions, which are easy to visualise and communicate to decision-makers and other stakeholders. In addition, very limited information is required from decision-makers and experts to construct SD, contrary to the traditional mean-variance analysis where the specification of a utility function is necessary \cite{Canessa2016, Markowitz2000}. An extension to the proposed approach is to introduce new uncertainty metrics by specifying order-preserving functions of the presented orderings.

\section{Data availability}
Dataset and code related to this article can be found at \url{https://github.com/vicvolodina93/OrderHeating/tree/master}, hosted at Github. 

\appendix
\label{sec:appendix}

\section{Scaling and dispersion orderings}
\label{subsec:scaling}
In our analysis, the variables produced by the model, i.e. NPC and $CO_2$-equivalent emissions, both have natural units of measurements, providing a natural scaling. We claim that such scalings are arbitrary, and therefore the choice of units should not affect the ordering. We investigate the effect of scaling on the $L_1$ and generalised dispersion orderings, respectively.

First, consider the $L_1$ distance ordering.  Here, the scaling of the variables impacts the relative contribution of each one in the calculation of the $L_{1}$ distance.  To illustrate, consider the case in which $d=2$ and therefore $L_{1}$ is given by
$$D(\bb{x}_1,\bb{x}_2)=|x_{11}-x_{21}|+|x_{12}-x_{22}|.$$
Suppose that $x_{12}$ and $x_{22}$ are on the range $[0, 1]$ whilst $x_{11}$ and $x_{21}$ are on the range $[0, \lambda]$ where $\lambda>1$. Define $x'_{11}$ and $x'_{21}$ to be normalised values of $x_{11}$ and $x_{21}$, scaled so that they are on the $[0, 1]$ range.  The $L_1$ distance can therefore be written as
\begin{align*}
D(\bb{x}_1, \bb{x}_2)&=|\lambda x'_{11}-\lambda x'_{21}|+|x_{12}-x_{22}|\\
&=\lambda|x'_{11}-x'_{21}|+|x_{12}-x_{22}|,
\end{align*}
From the expression above, it is clear that we place a higher weighting on the first term of the sum due to scaling by a factor of $\lambda$, and therefore these variables are not considered equally. As a result, we can conclude that the $L_1$ ordering is sensitive to scale.  Therefore, in this paper, we pre-process the data and standardise the variables to the range $[0, 1]$. 

We now consider the impact of scaling on the generalised simplex dispersion ordering. To illustrate, we consider the case in which $k=d=2$ and compute the area of a triangle formed with three distinct points $\bb{x}_1, \bb{x}_2$ and $\bb{x}_3$, i.e.
\begin{align*}
\mbox{vol}\{\triangle(\bb{x}_1, \bb{x}_2, \bb{x}_3)\}&=\frac{1}{2}
\begin{vmatrix}
x_{11} & x_{12} & 1\\
x_{21} & x_{22} & 1\\
x_{31} & x_{32} & 1
\end{vmatrix}\\
&=\frac{1}{2}\Bigg[x_{11}\begin{vmatrix} x_{22}&1\\
x_{32} &1
\end{vmatrix} -x_{21}\begin{vmatrix} x_{12}&1\\
x_{32} &1
\end{vmatrix}+x_{31}\begin{vmatrix} x_{12}&1\\
x_{22} &1
\end{vmatrix}\Bigg]
\end{align*}
where $x_{11}, x_{21}, x_{31}$ and $x_{12}, x_{22}, x_{32}$ are in the ranges $[0, \lambda]$ and $[0, 1]$ respectively. We define $x'_{11}, x'_{21}$ and $x'_{31}$ as the normalised values of $x_{11}, x_{21}$ and $x_{31}$ and re-write the expression for the area of a triangle as:
\begin{align*}
\mbox{vol}\{\triangle(\bb{x}_1, \bb{x}_2, \bb{x}_3)\}&=\frac{1}{2}
\begin{vmatrix}
\lambda x'_{11} & x_{12} & 1\\
\lambda x'_{21} & x_{22} & 1\\
\lambda x'_{31} & x_{32} & 1
\end{vmatrix}\\
&=\frac{1}{2}\Bigg[\lambda x'_{11}\begin{vmatrix} x_{22}&1\\
x_{32} &1
\end{vmatrix} -\lambda x'_{21}\begin{vmatrix} x_{12}&1\\
x_{32} &1
\end{vmatrix}+\lambda x'_{31}\begin{vmatrix} x_{12}&1\\
x_{22} &1
\end{vmatrix}\Bigg]\\
&=\frac{\lambda}{2}\Bigg[x'_{11}\begin{vmatrix} x_{22}&1\\
x_{32} &1
\end{vmatrix} -x'_{21}\begin{vmatrix} x_{12}&1\\
x_{32} &1
\end{vmatrix}+x'_{31}\begin{vmatrix} x_{12}&1\\
x_{22} &1
\end{vmatrix}\Bigg].
\end{align*}
The effect of scaling is therefore to multiply the area of the triangle by $\lambda$.  The re-scaling must therefore preserve the simplex ordering, and we conclude that the simplex volume is a scale-free, homogeneous measure.  We argue that this is a major advantage of the generalised simplex ordering since no arbitrary pre-processing of a data set is required. \par
\FloatBarrier

\bibliographystyle{apa}
\bibliography{references}
\end{document}